\documentclass[aps,pra, twocolumn, superscriptaddress,amssymb,showpacs]{revtex4-1}
\usepackage{graphicx}
\usepackage{epstopdf}

\begin{document}
\title{ Phase-locking at low-level of quanta}

\author{G. H. Hovsepyan}
\email[]{gor.hovsepyan@ysu.am}
\affiliation{Institute for Physical Researches,
National Academy of Sciences,\\Ashtarak-2, 0203, Ashtarak,
Armenia}

\author{A.~R.~Shahinyan}
\email[]{anna\_shahinyan@ysu.am}
\affiliation{Institute for Physical Researches,
National Academy of Sciences,\\Ashtarak-2, 0203, Ashtarak, Armenia}

\author{Lock Yue Chew}
\email[]{lockyue@ntu.edu.sg}
\affiliation{Nanyang Technological University,
21 Nanyang Link, SPMS-PAP-04-04, Singapore 637371}\affiliation{Complexity Institute, 
Nanyang Technological University, 18 Nanyang Drive, Singapore 637723}

\author{G.~Yu.~Kryuchkyan}
\email[]{kryuchkyan@ysu.am}
\affiliation{ Yerevan State University, Centre of Quantum Technologies and New Materials, Alex Manoogian 1, 0025,
Yerevan, Armenia}\affiliation{ Armenian State Pedagogical University after Kh. Abovyan, Tigran Mets ave. 17, 0010, Yerevan, Armenia}

\begin{abstract}
{We discuss phase-locking phenomena at low-level of quanta for parametrically driven nonlinear Kerr resonator (PDNR) in strong quantum regime. Oscillatory mode of PDNR is created in the process of a degenerate down-conversion of photons under interaction with a train of external Gaussian pulses. We calculate the Wigner functions of cavity mode showing two-fold symmetry in phase space and analyse formation of phase-locked states in the regular as well as the quantum chaotic regime.}
\end{abstract}

\maketitle

\section{Introduction}\label{introduction}

The parametric phase-locked oscillator is known to possess a wide-ranging set of applications in both the fundamental and the applied sciences. In recent years, this device has become a significant part in the experimental implementation of basic quantum optical systems and they are also envisaged to be a core component of quantum computers \cite{Lin,Monroe}. One of the standard system displaying phase-locking behavior is the optical parametric oscillator (OPO) which has been shown to be an efficient source of squeezed light \cite{wu}.
The subharmonic oscillatory mode of OPO excited through the degenerate down-conversion process is found to exhibit two phase stability. This indicates the presence of two stable states in the optical mode above the generation threshold, with the two states having equal photon number but different phases \cite{Drum,Drum1,Drum2}.
As a result, the Wigner function of the subharmonic mode acquires a two-fold symmetry in phase space.

Such phase-locking behavior can also be realized in the nondegenerate (double resonant) optical parametric oscillator (NOPO) by including additional intracavity quarter-waveplate to provide polarization mixing between two orthogonally polarized modes of the subharmonics \cite{mason1,Fabre,Long}. A full quantum mechanical treatment of this system has been presented in \cite{hadam,nadam} on the generation of continuous-variable entangled states of light beams under mode phase locked condition, while experimental realizations have been achieved in \cite{laura,laurb,feng}.
The experimental observation of dynamical signatures of self-phase-locking phenomenon in a triply resonant degenerate OPO was reported in \cite{zondya}. On the other hand, the cascaded phase-locked oscillators for the production of three-photon states were proposed in \cite{kryuchkyana,kryuchkyanb,zondyb,kryuchkyanc}.

Phase-locking transition was also observed in a chirped superconducting Josephson resonator \cite{Naaman}. More recently, the Josephson parametric phase-locked oscillator was demonstrated, and it was applied to detect binary signals with the identified digital information stored in the form of two oscillatory distinct phases \cite{Lin}. This enables the accurate measurement of the state of superconducting qubit without affecting the integrity of the stored information.

Another important implementation that exploits the phase-locking mechanism is the combination of OPO with the intracavity Kerr nonlinear element, the so-called parametrically driven nonlinear resonator (PDNR) \cite{PNO1,PNO11}. This system was first proposed as an optical parametric oscillator in the pulsed regime with the incorporation of an intracavity third-order nonlinear element leading to Kerr-interaction \cite{holm}. A complete quantum treatment of phase-locked PDNR has been developed in terms of the Fokker-Planck equation in complex P-representation \cite{a33,b33,Milb}. The quantum regime of PDNR requires a comparatively high level of third-order Kerr nonlinearity with respect to dissipation. In this regime, the  pulsed PDNR has been proposed for the production of quantum superposition states and two-photon Fock states \cite{ Tigran}.

Up till now, the phenomenon of phase-locking has been examined mainly from the mean-field approach which is fully justifiable in the case of relatively high photon number of cavity modes as well as for the regular operational regimes of parametric devices, for instance via the use of continuous-wave or pulsed driving fields. In this paper, we shall investigate into phase-locking at a low-level of excitation numbers of cavity mode; in the strong quantum regime; and in complete consideration of dissipation and decoherence. Moreover, we shall analyse into phase-locking within the chaotic regime of PDNR. In other words, we study into properties of dissipative chaos for oscillatory open systems which display phase-locking behavior.

The most successful approach of probing quantum dissipative chaos is based on quantum tomographic methods, which involve measurements of Wigner function in phase space. In consequence, quantum chaos can be detected through a comparison between the contour plots of the Wigner functions and the strange attractors that characterize dissipative chaos in classical Poincar\'e section \cite{PhysE,mpop,Leon}. Note that such analysis seems to be rather
qualitative than quantitative for strong quantum regime and for the range of low-level oscillatory excitation numbers, where the validity of semiclassical
equation is questionable. Thus, in this paper we provide only quantum analysis of chaos based on the Wigner functions without any analysis from the semiclassical approach.

Note that alternative approaches of probing quantum dissipative chaos involve consideration of entropic characteristics; analysis of statistics of excitation number \cite{K2,K3}; and application of methods based on fidelity decay \cite{Lloyd,Weins,L2}, Kullback-Leibler quantum divergence \cite{Kulback},
and the purity of quantum states \cite{purOur}.

Our paper is arranged as follow. In Sec. II, we provide a brief description of PDNR driven by a train of Gaussian pulses. In Sec. III, we study the phase-locking phenomena at a low-level of excitation number for both the regular and chaotic regimes of the PDNR. Finally, we summarize our results in Sec. IV.

\section{ PDNR under pulse train}\label{SecPD}

We consider a composite nonlinear one-mode resonator involving second-order and third-order Kerr nonlinearities excited by a train of pulses. The system is thus a parametrically driven Kerr resonator with the following Hamiltonian:
\begin{eqnarray}
H=\hbar \omega_{0}a^{+}a + \hbar \chi (a^{+}a)^{2}+~~~~~~~~~~~~~~~~~~~~~~\nonumber ~\\ \hbar f(t)({\Omega} e^{-i\omega t}a^{+2} + {\Omega^{*}}e^{i\omega t}a^{2})+H_{loss}\,.
\label{H}
\end{eqnarray}
Here, $a^{+}$ and $a$ are the oscillatory creation and annihilation operators, $ \omega_{0}$ is the oscillatory frequency, $ \omega$ is the mean frequency of the driving field, and $\chi$ is the strength of the nonlinearity which is proportional to the third-order susceptibility for the case of Kerr-media. The coupling constant $\Omega f(t)$ is proportional to the second-order susceptibility and the time-dependent amplitude of the driving field. It consists of Gaussian pulses with duration $T$ separated by time interval $\tau$ as follow:
\begin{equation}
f(t)=\sum_{n=0}^{\infty}{e^{-(t - t_{0} - n\tau)^{2}/T^{2}}}\,.
\label{driving}
\end{equation}
Note that $H_{loss}=a \Gamma^{+} + a^{+} \Gamma$ is responsible for the linear loss of the oscillatory mode due to coupling to a heat reservoir which gives rise to a damping rate of $\gamma$.

The reduced density operator of the oscillatory mode $\rho$ obeys the transformation $\rho \rightarrow e^{-i(\omega /2)a^{+}at} \rho e^{i(\omega /2)a^{+}at}$ in the interaction picture. It is governed by the following master equation:
\begin{eqnarray}
\frac{d \rho}{dt} =-\frac{i}{\hbar}[H_{0}+H_{int}, \rho] + ~~~~~~~~~~~~~~~~~~~~~~~~\nonumber ~\\
\sum_{i=1,2}\left( L_{i}\rho
L_{i}^{+}-\frac{1}{2}L_{i}^{+}L_{i}\rho-\frac{1}{2}\rho L_{i}^{+}
L_{i}\right)\label{master}
\end{eqnarray}
within the framework of the rotating-wave approximation. Note that $L_{1}=\sqrt{(N+1)\gamma}a$ and $L_{2}=\sqrt{N\gamma}a^+$ are the Lindblad operators, $\gamma$ is the dissipation rate, and $N$ denotes the mean number of quanta of the heat bath. Furthermore,
\begin{eqnarray}
H_{0}=\hbar \Delta a^{+}a\,, ~~~~~~~~~~~~~~~~~~~~~~~\nonumber ~\\
H_{int}= \hbar \chi (a^{+}a)^{2} + \hbar f(t)({\Omega}a^{+2} +
{\Omega^{*}}a^{2})\,,\label{hamiltonian4}
\end{eqnarray}
with $\Delta=\omega_{0}-\omega/2$ being the detuning between the half frequency of the driving field $ \omega/2$ and the oscillatory frequency $\omega_{0}$. To study pure quantum effects, we focus on cases of very low reservoir temperature for which the mean number of reservoir photons $N=0$.

We consider our system to be in the regime of strong Kerr nonlinearities with respect to dissipation, i.e. $\chi/\gamma>1$. Recent progress in circuit QED, superconducting systems and solid-state artificial atoms has opened up new avenues for the design of device configurations based on our model. In particular, the Hamiltonian given by Eq. (\ref{H}) with $f(t)=1$, describes a Josephson junction embedded in a transmission-line resonator with the effect of the quadratic part of the Josephson potential being taken into account exactly. In this case, the raising ($a^{+}$) and lowering ($a$) operators describe the normal mode of the resonator plus junction circuit. The nonlinearity is found to lead to self-Kerr effects. By replacing the single junction with a SQUID, the Kerr coefficient then allows the parametric term to describe a degenerate two-photon excitation by microwave light \cite{Bour}.

We solve the master equation given by Eq. (\ref{master}) numerically based on the method of quantum state diffusion (QSD) \cite{qsd}. The application of this technique to the studies of driven nonlinear oscillators and OPOs can be gleaned from the Refs. \cite{PhysE, mpop, K2, K3}.

 In the semiclassical approach, the corresponding equation of motion for the dimensionless amplitude of oscillatory mode takes the following form:
\begin{equation}
\frac{d\alpha}{dt}= -i[\Delta + \chi + 2|\alpha|^2\chi]\alpha + if(t)\Omega\alpha^{*} -\gamma\alpha.
\label{semclass}
\end{equation}
This equation modifies the standard equation for parametric oscillator with Kerr nonlinearity for the case of pulsed excitation.

To proceed with further analysis we shall present at first 
the semiclassical steady-state solutions and stability properties
of PDNR  for the monochromatic excitation, $f(t)=1$ \cite{a33,b33}.
 In the standard analysis  the amplitude of the oscillatory mode is represented as $\alpha =
n^{1/2}exp(i \varphi)$ in term  of the intensity $n$ (in photon
number units) and the phase of the mode $\varphi$. We also assume the amplitude $\Omega=I^{1/2}exp(i\Phi)$, where  $I$ is the intensity and $\Phi$ is the
phase of the driving field. In this case above
threshold regime takes place for $I >
I_{th}=\frac{\gamma^{2}}{\Omega^{2}}(1+\frac{\Delta^{2}}{\gamma^{2}})$ and above-threshold solutions are
determined by the following expressions:

\begin{eqnarray}
 n=\frac{\gamma}{2 \chi} [\frac{\Delta}{\gamma}+(J -1)^{1/2}], \nonumber ~\\
\sin(\Phi -2\varphi)=J ^{-1/2},\label{hamiltonian}
\end{eqnarray}
where $J = \frac{ \Omega ^{2}}{\gamma^{2}}I$. 
 These results are obtained in over transient regime and for large
oscillatory mean excitation numbers, $n>>1$. Using these equations we conclude that the system displays regular behavior
of photon number  $n$ versus $J$ for positive detunings, otherwise, for the negative detunings, the bistable regime is realized. Beside this in the regular, above-threshold regime there exist actually
two stable steady states, which have equal intensities but opposite phases
\begin{equation}
\varphi=\frac{1}{2} \Phi \pm\pi m, \label{PhaseSol}
\end{equation}
  thus we observe a phase locking with two
fold symmetry.

It is well known that the phase-locking phenomenon reflects the two-fold symmetry of the Wigner function. It has been demonstrated in steady-state  regime of OPO  under monochromatic driving \cite{kryuchkyana,kryuchkyanb,kryuchkyanc}.  The situation of phase-locking with two-fold symmetry also arises for the composite system under consideration due to the quadratic form of the nonlinear term in the Hamiltonian as it is realized for the pulsed PDNR (\ref{H}). Indeed,  considering the transformations:
\begin{equation}
H^{\prime}=U^{-1}HU,~~ \rho^{\prime}=U^{-1}\rho U
\label{Transform}
\end{equation}
with the unitary operator
\begin{equation}
U=exp\left(i \theta a^{+}a \right) \label{TransformOp}
\end{equation}
we verify that the Hamiltonian (\ref{H}) and the density operator of
oscillatory mode satisfy the following commutation relations
\begin{equation}
\left[H,U\right]=0, \label{ComRel}
\end{equation}
\begin{equation}
\left[\rho(t),U\right]=0, \label{DisEf}
\end{equation}
if the parameters $\theta$ is chosen as $\theta =\pi$. On the whole, it is easy to demonstrate that  the Wigner function displays two-fold symmetry in its rotation around the origin of the phase space:
\begin{equation}
W\left(r,\theta+\pi\right)=W\left(r,\theta \right)\,.
\label{WigPol}
\end{equation}
Note that the polar coordinates $r,~\theta$ are related to the complex plane $X=\left(\alpha + \alpha^{*}\right)/2=r\cos \theta$ and $Y=\left(\alpha - \alpha^{*}\right)/2i=r\sin \theta$, with $\alpha$ corresponding to the field operator $a$ in the positive P-representation. This result is obtained in the general form for arbitrary time-dependent amplitudes of external field and for all operational regimes of PDNR. In fact, such two-fold symmetry persists in the Wigner functions when the PDNR is taken into the quantum chaotic regime, which will be discussed in the next section.

In the following, we consider the phase-locking phenomena based on the Wigner functions in phase space within both the regular and chaotic regimes of the PDNR. The Wigner function is calculated by averaging an ensemble of quantum trajectories for a definite time instant. By analyzing the operational regimes of PDNR driven by the pulse train, we have concluded that the regular regime is mainly realized for positive detunings $\Delta> 0$, while chaotic dynamics takes place for negative detunings $\Delta< 0$. This result is in accordance with the above discussion of the case of monocromatic excitation of PDNR that   exhibits bistability
for a negative detung. Indeed, just in the  case  of negative detuning the contribution of pulse train leads to controlling transition from bistable to chaotic dynamics  in
terms of the semiclassical solution.  In addition, for chaos to happen, the other parameters of our model have to satisfy the following criteria:
 $\Omega \simeq \vert\Delta \vert$, and $\pi/2 \leq\tau/T \leq 2 \pi$. It should be noted that the above conditions are specific to the pulsed PDNR, since for the standard pulse driven anharmonic oscillator without parametric term, quantum chaotic regimes are realized for both signs of the detuning.
The typical results for the Wigner functions corresponding to the cases of positive and negative detunings for the pulse driven PDNR are illustrated in Figs. \ref{fig.minmax}, \ref{fig.wigners2} and \ref{fig.wigners1}.

\section{Phase-locking in order-to-chaos transition }

In this section, we analyse the phase-locking phenomena by determining the time-dependent excitation number and Wigner function of the oscillatory mode. We calculate these quantities by averaging over an ensemble of quantum trajectories for definite time instants, and with respect to the parameters $\Delta/\gamma $, $\chi/\gamma$, $\Omega/\gamma$ as well as the parameters of the pulses corresponding to the regular and chaotic regimes. Note that both the ensemble-averaged mean oscillatory excitation number and the Wigner function are nonstationary. They exhibit a periodic time-dependent behavior, i.e., they follow the periodicity of the driving pulses beyond the transient state of the system.

\begin{figure}
\includegraphics[width=8.6cm]{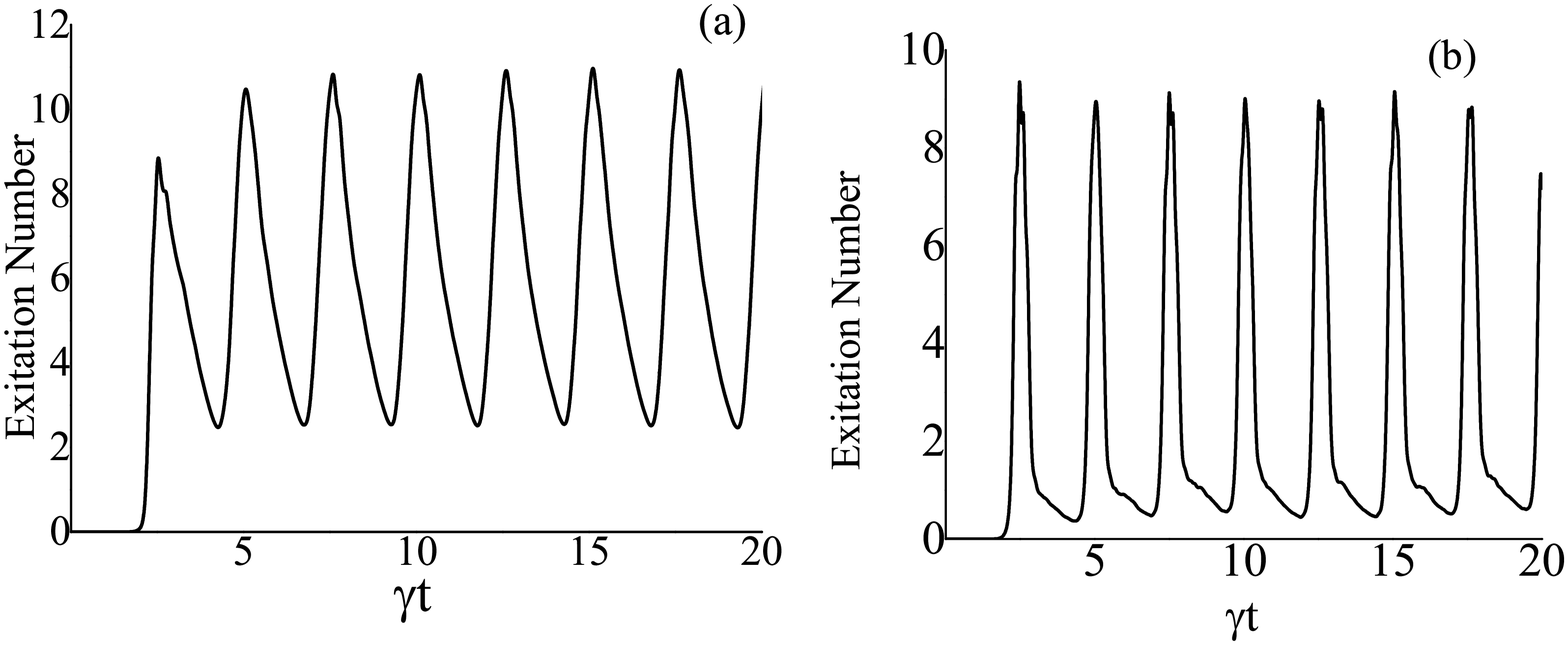}
\caption{Excitation number against time intervals for the negative and positive detuning. The parameters used are: $\chi/\gamma=1$, $\Omega/\gamma=20$, $T=0.5\gamma^{-1}$, $\tau=4\pi\gamma^{-1}/5$;  (a) $\Delta/\gamma = -20$, (b) $\Delta/\gamma = 20$.}
\label{fig.wigners}
\end{figure}

The typical results on the dynamics of the photon number are presented in Fig. \ref{fig.wigners} for the negative and positive value of the detuning corresponding to the (a) chaotic and (b) regular regimes. We observe that the system operates in the strong quantum regime at a level of small excitation number for these parameter values. The ensemble averaged mean excitation number is clearly regular in both regimes. It demonstrates the well known result that quantum dissipative chaotic dynamics is not evident in the dynamics of the mean oscillatory number. Nonetheless, it is possible to detect quantum chaotic behavior from other physical quantities, in particular, the Wigner function.

\begin{figure}
\includegraphics[width=8.6cm]{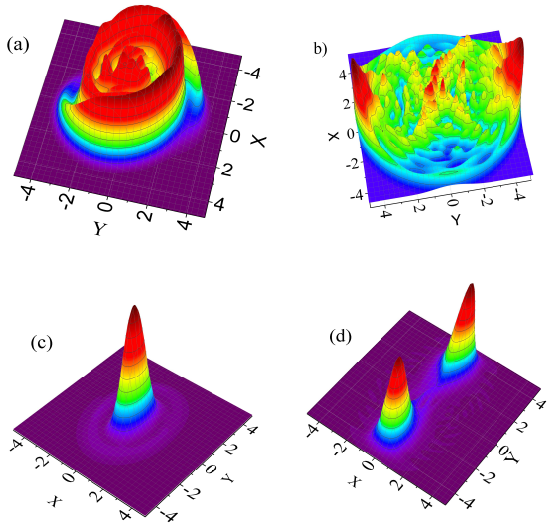}
\caption{The Wigner functions for the positive and negative detunings at time instants corresponding to the minimum (a), (c) and the maximum (b), (d) of the excitation number observed in Fig. \ref{fig.wigners}. The parameters employed are: (a), (b) $\Delta/\gamma = -20$, $\chi/\gamma=1$, $\Omega/\gamma=20$, $T=0.5\gamma^{-1}$, and $\tau=4\pi\gamma^{-1}/5$; (c), (d) $\Delta/\gamma = 20$, $\chi/\gamma=1$, $\Omega/\gamma=20$, $T=0.5\gamma^{-1}$, and $\tau=4\pi/5\gamma^{-1}$. Note that (a) and
(c) relate to the minimal value of the photon number, while (b) and (d) to the maximal value.  The Wigner functions of (a) and (b) are for the chaotic regime, while (c) and (d) the regular regime of PDNR.}
\label{fig.minmax}
\end{figure}

In Fig. \ref{fig.minmax}, we present our results on the Wigner function in the strong quantum regime for both the regular and chaotic behavior of the PDNR. Note that the results were selected at definite moment in time that corresponds to the minimal and maximal value of the mean excitation number depicted in Fig. \ref{fig.wigners}.
We observe that all the Wigner functions display two-fold symmetry in phase
space according to Eq. (\ref{WigPol}). In particular, Figs. \ref{fig.minmax}(c)
and \ref{fig.minmax}(d)  describe the formation of phase locked states for the
regular regime as the photon number of the oscillatory mode increases. The
single-peaked Wigner function which occurs at the minimum of the excitation
number $n_{min}$= 1, is squeezed in phase space (see Fig. \ref{fig.minmax}(c)) indicating the
formation of squeezed states in the oscillatory mode. By increasing the
excitation number, we observe the formation of two squeezed humps (see Fig.
\ref{fig.minmax}(d)) at the maximum of the excitation number $n_{max}$= 9. The two humps
correspond to two states of equal photon number, but with two different phases
of the cavity mode of the PDNR, which is above the generation threshold of the
semiclassical approach. We notice that the distance between the two humps
depends on the excitation number of the optical mode. Interestingly, our
results have uncovered the occurrence of phase-locking behavior at relatively
small excitation number in the strong quantum regime.

Our results for the case of negative detuning are depicted in Figs. \ref{fig.minmax}(a) and \ref{fig.minmax}(b) for time instants that correspond to the minimal and maximal values of the excitation number (see Fig.1 (a)). The latter result is cardinally different from the Wigner function obtained for the regular operational regime. While the contour plots of the Wigner function for regular dynamics are clearly bell-shaped and localized in a narrow region of phase space, the phase space distribution for the case of negative detuning is observed to be wider. In fact, we also observe a broadening of the corresponding excitation number distribution $P(n)$. The shape of the distribution is found to change irregularly depending on the duration $T$ and time interval $\tau$ between pulses. We conjecture that these Wigner functions depict the chaotic behaviour of PDNR in the strong quantum regime. A detailed analysis of quantum chaos based on a comparison between the contour plots of the Wigner function and the corresponding classical Poincar\'e section has been performed for the standard nonlinear resonator driven by a train of periodic pulses \cite{purOur2}. It was demonstrated in \cite{purOur2} that for comparatively small values of the ratio
$\chi/\gamma$, the contour plots of Wigner functions are relatively similar to the strange attractors in the Poincar\'e section. The main difference being the existence of fine fractal structures within the Poincar\'e section, which are absent in the Wigner function. This is due to the Hiesenberg uncertainty principle which prevents sub-Plankian structures to appear in phase space. The consequence is a loss of correspondence between the quantum and classical distributions in the deep quantum regime.

By analysing the behavior of phase-locking in the chaotic regime of PDNR, we
can easily conclude that the Wigner function exhibits two-fold symmetry under a
rotation of angle $\pi$ around its origin in phase space based on the general
formula given by Eq. (\ref{WigPol}). Thus, the parametric interaction of cavity
mode in PDNR is observed to display two-fold symmetry in the phase space for
both operational regimes. The Wigner function in Fig. \ref{fig.minmax}(a) has the form of a
broken one-peaked localized state, while the two maxima at the Wigner function
of Fig. \ref{fig.minmax}(b) reflect the track of phase-locked states in the
quantum chaotic regime.

\begin{figure}
\includegraphics[width=8.6cm]{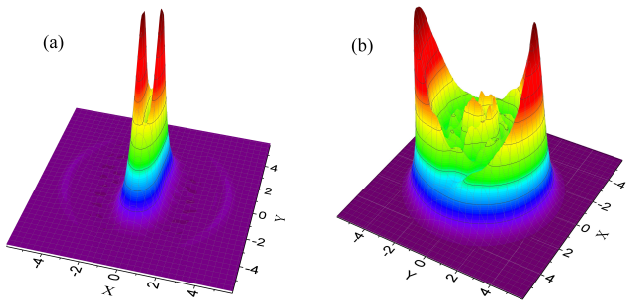}
\caption{The Wigner function for the case of positive and negative detuning, and for a time instant of mean excitation number that lies between the minimal and maximal values. The parameters used are: $\chi/\gamma=1$, $\Omega/\gamma=20$, $T=0.5\gamma^{-1}$, $\tau=4\pi\gamma^{-1}/5$; (a) $\Delta/\gamma = 20$, (b) $\Delta/\gamma = -20$.}
\label{fig.wigners2}
\end{figure}

It is also important to examine the phase-locking behavior by considering the system time-evolution. To achieve this goal, we have calculated the Wigner function for time instants that lie between the minimal and maximal value of the excitation number. The results depicted in Fig. \ref{fig.wigners2} show the  initial stage of phase splitting of the optical mode for the two phases in both the regular and chaotic regimes.

\begin{figure}
\includegraphics[width=8.6cm]{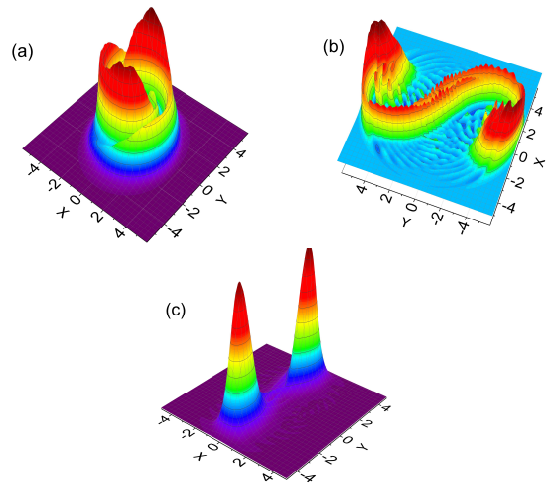}
\caption{(Color online) The Wigner functions for negative detuning for time instants that correspond to the minimal (a) and the maximal (b) values of the mean excitation number. The parameters used are: $\Delta/\gamma = -7.5$, $\chi/\gamma=0.5$, $\Omega/\gamma=10$, $T=0.25\gamma^{-1}$, and $\tau=2 \pi/5\gamma^{-1}$. The Wigner function for the regular regime (c), with time instant at the maximal value of the mean excitation number. The parameters employed are: $\Delta/\gamma = 15$, $\chi/\gamma=1$, $\Omega/\gamma=10$, $T=0.5\gamma^{-1}$, and $\tau=2\pi\gamma^{-1}$.}
\label{fig.wigners1}
\end{figure}

We found that the above results are typical for PDNR in strong quantum regime. In Fig. \ref{fig.wigners1}, we have presented additional Wigner functions at other parameter values of $\Delta/\gamma$, $\chi/\gamma$ and $\Omega/\gamma$. In particular, we observe that a decrease in the amplitude of the driving field would lead to a decrease in the distance between the humps of the two localized states. This is apparant by comparing the results of Fig. \ref{fig.minmax}(d) with Fig. \ref{fig.wigners1}(c) for the regular regime. In the case of quantum chaotic regimes (see Figs. \ref{fig.wigners1}(a) and (b)), we notice that a slight change in the system parameters can lead to subtle variations in the Wigner function. Indeed, this is easily observed in Figs. \ref{fig.wigners1}(a) and (b), where a smaller nonlinearity and field amplitude have been used relative to those employed in Fig. \ref{fig.minmax}.

\section{Summary}

We have demonstrated that the phase-locking phenomenon, which is predicted by the mean-field approximation to arise under the situation of a relatively high photon number in the optical modes, can also occur at the level of a few quanta. We have performed the analysis based on a nonlinear Kerr oscillator parametrically driven by a train of pulses in the strong quantum regime. The analysis is carried out by calculating the Wigner functions of cavity mode for definite time instants with temporal pulses of different widths in both the regular and chaotic regimes. Our results have shown that the Kerr resonator parametrically driven by a train of Gaussian pulses exhibits specific chaotic behavior that is very different from those displayed by the standard nonlinear Kerr resonator in the pulsed regime without parametric excitations. Furthermore, in contrast to the case of standard nonlinear Kerr resonator, the quantum chaotic behavior of the PDNR is found to take place only for negative $\Delta<0$ detunings. More importantly, we have uncovered features of phase-locking effects from this quantum dissipative chaotic system through the analysis of its Wigner functions.

\begin{acknowledgments}
We acknowledge the support from the Armenian State Committee of Science, the Project No.13-1C031.  G. Yu. Kryuchkyan thanks the Nanyang Technological University for hospitality.
\end{acknowledgments}


\begin{thebibliography}{0}

\bibitem{Lin} Z.R. Lin, K. Inomata, K. Koshino, W. D. Oliver, Y. Nakamura, J. S. Tsai, and T. Yamamoto, Nature Communications
{\bf5}, 4480 (2014).

\bibitem{Monroe} R. Islam, W. C. Campbell, T. Choi, S. M. Clark, C. W. S. Conover, S. Debnath, E. E. Edwards, B. Fields, D. Hayes, D. Hucul, I. V. Inlek, K. G. Johnson, S. Korenblit, A. Lee, K. W. Lee, T. A. Manning, D. N. Matsukevich, J. Mizrahi, Q. Quraishi, C. Senko, J. Smith, and C. Monroe, Optics Letters, {\bf39},3238 (2014).

\bibitem{wu} L. A. Wu,  H. J. Kimble, J. L. Hall, and  H.Wu, Phys. Rev. Lett. {\bf 57}, 2520 (1986).

\bibitem{Drum} P. Kinsler and P. D. Drummond, Phys. Rev. A {\bf 43}, 6194
(1991).

\bibitem{Drum1} P. D. Drummond and P. Kinsler, Quantum Semiclassic. Opt. {\bf 7}, 727 (1995).

\bibitem{Drum2} P. Kinsler and P. D. Drummond, Phys. Rev. A {\bf 52}, 783 (1995).

\bibitem{mason1} Mason E. I. and Wong N.C., Opt. Lett. {\bf23}, 1733 (1998)

\bibitem{Fabre}Fabre C., Mason E.I., and Wong N.C., Opt. Commun. {\bf170}, 299 (1999)

\bibitem{Long} Longchambon L. et al., Eur. Phys. J. D {\bf30}, 279 (2004)

\bibitem{hadam}H. H. Adamyan and G. Yu. Kryuchkyan, Phys. Rev. A {\bf69}, 053814 (2004)

\bibitem{nadam}H. H. Adamyan, N. H. Adamyan, S. B. Manvelyan and G. Yu. Kryuchkyan, Phys. Rev. A {\bf73}, 033810 (2006)

\bibitem{laura}J. Laurat, T. Coudreau, G. Keller, N. Treps, and C. Fabre, Phys. Rev. A {\bf70}, 042315 (2004)

\bibitem{laurb}J. Laurat, T. Coudreau, L. Longchambon, and C. Fabre, Opt. Lett. {\bf30}, 1177 (2005)

\bibitem{feng}S. Feng and O. Pfister, Phys. Rev. Lett. {\bf92}, 203601 (2004)

\bibitem{zondya}J-J. Zondy, D. Kolker, and F. N. C. Wong, Phys. Rev. Lett. {\bf93}, 043902 (2004)

\bibitem{kryuchkyana}G.Yu. Kryuchkyan, N.T. Muradyan, Phys. Lett. {\bf286},113–120 (2001)

\bibitem{kryuchkyanb}G. Yu. Kryuchkyan, L. A. Manukyan and N. T. Muradyan, Opt.Commun. {\bf190}, 245 (2001)

\bibitem{zondyb}J. J. Zondy, A. Tallet, E. Ressayre, and M. LeBerre, Phys. Rev. A {\bf63}, 023814 (2001)

\bibitem{kryuchkyanc}D.A. Antonosyan, T.V. Gevorgyan,G.Yu. Kryuchkyan, Phys. Rev. A {\bf83}, 043807 (2011)

\bibitem{Naaman} O. Naaman, J. Aumentado,L. Friedland, J. S. Wurtele, and I. Siddiqi, Phys.Rev. Letters, {\bf101},  117005 (2008)

\bibitem{PNO1} B. Wielinga and G. J. Milburn, Phys. Rev. {\bf A 48}, 2494  (1993).

\bibitem{PNO11} B. Wielinga and G. J. Milburn, Phys. Rev. {\bf A 49}, 5042
(1994).

\bibitem{holm} G. J. Milburn, C. A. Holmes, Phys.Rev.  {\bf A 44}, 4704 (1991).

\bibitem {a33} G. Yu. Kryuchkyan and K. V. Kheruntsyan, Opt.
Comm. {\bf 120}, 132 (1996).

\bibitem {b33} K. V. Kheruntsyan, D. S. Krahmer, G. Yu. Kryuchkyan, K. G.Petrossian, Opt.  Comm. {\bf 139}, 157 (1997).

\bibitem{Milb} C. H. Meaney, H. Nha, T. Duty and G. J. Milburn, EPJ Quantum
Technology, {\bf 1}, 7 (2014).

\bibitem{Tigran} T. V. Gevorgyan and G. Yu. Kryuchkyan, Journal of Modern Optics {\bf 60}, 860 (2013).

 \bibitem {PhysE} H. H. Adamyan, S. B. Manvelyan and G. Yu. Kryuchkyan, Phys. Rev. E, 64, 046219, (2001).

\bibitem{mpop} T. V. Gevorgyan, S. B. Manvelyan, A. R. Shahinyan, G. Yu. Kryuchkyan, Dissipative Chaos in Quantum Distributions. In: Modern
Optics and Photonics: Atoms and Structured Media. Eds: G. Kryuchkyan, G.
Gurzadyan and A. Papoyan, World Scientific, 2010.

\bibitem {Leon} A. Kowalewska-Kud{\l}aszyk, J.K. Kalaga, W. Le\'{o}nski, Phys. Rev. E {\bf 78}, 066219 (2008).

\bibitem {K2} G. Yu. Kryuchkyan and S. B. Manvelyan, Phys. Rev. Lett., {\bf 88}, 094101, (2002).

\bibitem {K3} G. Yu. Kryuchkyan and S. B. Manvelyan, Phys. Rev. A, {\bf 68}, 013823 (2003).

\bibitem {Lloyd} J. Emerson, Y. S. Weinstein, S. Lloyd and D. G. Cory. Phys. Rev. Lett., {\bf 89}, 284102-1, (2002).

\bibitem {Weins} Y. S. Weinstein, S. Lloyd and C. Tsallis. Phys. Rev. Lett., {\bf 89}, 214101-1, (2002).

\bibitem{L2} A. Kowalewska-Kud{\l}aszyk, J.K. Kalaga, W. Le\'{o}nski, Phys.
Lett., A 373, 1334 (2009).

\bibitem {Kulback}  A. Kowalewska-Kud{\l}aszyk, J. K. Kalaga, W. Le\'{o}nski and V. Cao Long, Phys. Lett.,  A {\bf 376}, 1280 (2012).

\bibitem {purOur} A. R. Shahinyan, Lock Yue Chew, G. Yu. Kryuchkyan, Phys. Let. A {\bf 377}, 2743, (2013).

\bibitem{Bour} J. Bourassa, F. Beaudoin, Jay M. Gambetta, and A. Blais, Phys. Rev. A 86, 013814 (2012).

\bibitem {qsd} I. C. Percival, Quantum State Diffusion, Cambridge University Press, Cambridge, (2000).

\bibitem {purOur2} T. V. Gevorgyan, A. R. Shahinyan, Lock Yue Chew, and  G. Yu. Kryuchkyan, Phys. Rev. E {\bf 88}, 022910, (2013).

\end{thebibliography}
\end{document}